# CAPODAZ: A containerised authorisation and policy-driven architecture using microservices

Dimitrios Kallergis [a,*], Zacharenia Garofalaki [a,b], Georgios Katsikogiannis [b], Christos Douligeris [b]

[a] *Department of Informatics & Computer Engineering, University of West Attica, Greece*
[b] *Department of Informatics, University of Piraeus, Greece*

a b s t r a c t

The microservices architectural approach has important benefits regarding the agile applications' development and the delivery of complex solutions. However, to convey the information and to share the data amongst services in a verifiable and stateless way, there is a need to enable appropriate access control methods and authorisations. In this paper, we study the use of policy-driven authorisations with independent fine-grained microservices in the case of a real-world machine-to-machine (M2M) scenario using a hybrid cloud-based infrastructure and Internet of Things (IoT) services. We also model the authentication flows which facilitate the message exchanges between the involved entities, and we propose a containerised authorisation and policy-driven architecture (CAPODAZ) using the microservices paradigm. The proposed architecture implements a policy-based management framework and integrates in an on-going work regarding a Cloud-IoT intelligent transportation service. For the in-depth quantitative evaluation, we treat multiple distributions of users' populations and assess the proposed architecture against other similar microservices. The numerical results based on the experimental data show that there exists significant performance preponderance in terms of latency, throughput and successful requests.

## 1. Introduction

In resource-constrained communication paradigms, machine-to-machine (M2M) communications play a significant role in the service delivery [1]. To facilitate the real-time monitoring of the condition information, device data collection, reporting, remote maintenance, and telemetry, we need to enable the secure transmission of the context and the authenticated information. In the application domain, the authentication enables the base stations to corroborate the sensory data from the M2M entities and to control the authorised access to the M2M application systems.

When one tries to extend the M2M communications to the Internet of Things (IoT) ecosystem by also incorporating cloud-based infrastructures, there is a great need to efficiently and effectively meet the new security requirements that appear and to address the appropriate access control services needed for identification and privacy protection [2,3]. In the context of M2M-IoT security, various issues have been explored: the resource-constrained devices and their distributed nature, the requirements of application and network gateways interconnection, and the multiple load patterns in heterogeneous application domains [4].

Beyond any doubt, greater attention should be paid to ensure that there exist appropriate security controls and secure service levels across various sectors, including cross-border, inter-domain, and industrial. Due to the growing concerns in access management techniques and the numerous access controls which have been proposed recently [4–6], as well as because of the emerging Cloud-IoT era, there is a widespread interest in expanding policy-driven decision-based systems and in defining adaptive policy evolution, effective interoperability, and efficient performance mechanisms in such environments [7–9]. On these grounds, Kavin and Ganapathy [10] introduce a privacy-preserving model for Cloud- and IoT-based applications using the Chinese Remainder Theorem, and the authors in [11,12] propose a secure architecture of IoT- and fog-enabled systems which leverages cryptographic methods to comply with the data minimisation principle included in the EU General Data Protection Regulation (GDPR)[1].

* Corresponding author.
  *E-mail addresses:* d.kallergis@uniwa.gr (D. Kallergis), z.garofalaki@uniwa.gr (Z. Garofalaki), gkatsikog@unipi.gr (G. Katsikogiannis), cdoulig@unipi.gr (C. Douligeris).

---

[1] https://gdpr.eu/

Regarding performance and interoperability, one of the contemporary trends in developing Cloud-IoT services is the microservice architecture. Microservices allow for the development of creative application architectures and delivery solutions. The microservice architecture, by dividing systems and applications into smaller and independent elements, is a more granular way of implementing a service-oriented architecture (SOA) and it has become a critical building block in the deployment of Cloud-IoT applications [13–15]. The microservices can adapt rapidly to the operational challenges and facilitate more responsive and nimble software applications. To tackle the issues of service integration, delivery, and repartitioning, there exist several approaches that incorporate Docker containers [15–17]. Containers, by default, share the same operating system kernel and are often described as lightweight virtual machines. They substantially reduce the resource overhead and, thus, they can be ideally utilised as a building block in multi-agent solutions on dynamically changing environments.

Preuveneers and Joosen [18] discuss the authorisation enforcement and claim that, when delegating a policy, a trade-off exists between the latency overhead and the performance. They propose the development of a microservices framework using the Spring Cloud, the Netflix Hystrix library, and ForgeRock's OpenAM deployed in a Docker container. However, there is no mechanism for controlling and evaluating the access delegations given the complexity of the authorisation policies and the computational characteristics of the solution.

As mentioned, even though several authorisation schemes for Cloud-IoT environments already exist in the literature, the existing research work does not consider the implementation of policy-driven authorisations on containerised microservices. To tackle this issue and to efficiently manage the massive and bursting number of Cloud-IoT actors, we need an architecture that supports policy-driven M2M-IoT and dynamic authorisations for microservices by leveraging capability-based exchanges to gain access to the resources. By managing the access policies and enforcing the business rules accurately and efficiently, the effectiveness of the policy-decision service solutions for M2M-IoT entities is expected to improve. Furthermore, such an architecture needs to also address concerns about time consumption, scalability and elasticity. Vandikas and Tsiatsis [15] experiment on a cloud-based IoT infrastructure using microservices frameworks and evaluate its performance in terms of memory and binary footprint, as well as throughput. However, the authors do not consider either latency measurements or a containerised approach in their work.

In this paper, we consider Cloud-IoT authorisation control services, which can be comprised of several distributed (multi-agent) software subcomponents. On these grounds, we propose an innovative containerised authorisation and policy-driven architecture (CAPODAZ) by implementing a secure policy-based framework for M2M communications as proposed in our earlier work [19], and by integrating a hybrid cloud infrastructure [20]. Moreover, we utilise a real-world IoT service prototype [21,22] as a testbed and incorporate a cloud-testing service [23] to monitor performance metrics. In more detail, we leverage a microservice schema to control and manage the M2M-IoT actors, to trigger the required actions and to improve the service efficiency by maintaining stateless authentication assertions. This schema is delivered by multiple clusters of Docker containers to decrease the resource overhead and to maximise performance. Finally, this paper revisits the methodological approach presented in [15] by extending its scope of measurements and by implementing a containerised architectural approach. In this vein, we utilise and compare various RESTful microservices setups, namely, *Spark, Wildfly-Swarm* and *Spring-boot-Tomcat* which have small start-up times and, hence, have the potential to offer a faster service delivery.

The **major contributions** of this work are the following:

(i) We propose an innovative architecture (CAPODAZ) which facilitates access control and management of IoT actors by using the microservices paradigm. The proposed architecture provides policy-driven authorisations using access resource tokens in order to enable an adaptive schema.
(ii) We incorporate Docker-container clusters in a cloud-based infrastructure to reduce the associated resource overhead and maximise performance.
(iii) We evaluate the proposed architecture on a real-life Cloud-IoT scenario and compare it with similar microservice approaches.
(iv) We provide compelling evidence that the proposed architecture can be applied in a variety of Cloud-IoT scenarios that support capability-based exchanges to gain resources access.

The remainder of this paper is organised as follows. Section 2 describes our motivation and depicts the problem statement. Section 3 illustrates, analyses, and discusses the proposed architecture, while Section 4 evaluates the architecture in a real-life scenario. Section 5 concludes the paper and highlights likely the future work.

## 2. Motivation

In the following sections, we give the necessary information background on Cloud-IoT policy-aware engines and policy-driven authorisations to justify the implementation of the proposed architecture in the context of earlier works. Moreover, we shed light on capability-based authorisation tokens because of their efficiency when used in M2M-IoT solutions.

### 2.1. Policy-aware engines for Cloud-IoT management

A wide body of research has been carried out in the context of policy management for cloud services. Chhetri et al. [24] consider an adaptive management of service provisioning, which is realised by policy-aware software agents to ensure the appropriate service levels. The authors propose a framework which offers a quality-aware service by providing a decision-making mechanism based on the Service Level Agreements (SLA) negotiations. Besides, in the case of stakeholder-specific policies [25], the authors propose to dynamically decide which requests and tasks can be executed in a multi-tenant Platform-as-a-Service (PaaS) cloud model. They also claim that these tasks can be controlled by policy engines which support different types of domain- and application-specific policies.

The growing interest in mobile- and cloud-based interactions has triggered considerable research work. In [26], the authors propose sticky policies to manage the associated security concerns. Moreover, the authors point out the need to enhance the security and reliability of the cloud-based applications and to establish stronger relations between the policies and the service. In this context, the authors in [27], and [28] introduce a policy match maker and a reasoning engine to be used in cloud computing environments. They also propose that a policy engine which processes a request according to existing policy packages and customized parameters. They conclude that it is crucial to enable a syntactic and semantic analysis of the security requests for policy matching and management. In this context, Bhatt et al. [29] propose an attribute-based access control (ABAC) extension utilising user attributes, while Ferraiolo et al. [30] propose an access control framework architecture which reduces the host machine overhead, decreases administrative procedures and imposes confinement constrains that prevent unauthorised access.

## 2.2. Policy-driven authorisation management

Whilst considerable attention has been devoted to the development of access control and policy management specifications, numerous challenges exist on how to orchestrate the components across the services. There is still a great need to optimise the workflows and to considerably reduce the service load time. In [31], the authors present the orchestration of various authentication and authorisation processes with formal policy-based methods aiming to provide a secure access to the resources. They also describe the architectural concerns about how to facilitate the development, the evaluation, the enforcement and the management of the respective access control policies. Ngo et al. [4] propose an access control model for Intercloud multi-tenant scenarios by leveraging the exchanging tokens approach. Their approach is shown to have good performance in terms of the supported cloud services and client numbers. In the context of providing dynamic authorisation, Cirani et al. [32] propose an architecture targeting HTTP/CoAP (hyper-text transfer protocol/constrained application protocol) services to provide an authorisation framework, which can be integrated by invoking an external OAuth-based authorisation service (OAS). She et al. [33] present the evaluation of policies during service compositions. They argue that there is a need to depict the existence of a minimum policy model needs in order to be able to integrate various access control constraints with policy-based compliant services. Furthermore, Sinjilawi et al. [34] first propose the use of multi-level security classes of information flows in cloud-based solutions, where there is a need to identify and present the access control limitations, and, then, they propose a unified access control model with integrated components.

Policy-driven authorisation in Cloud- and IoT-based systems is widely attractive due to the need of complying in private and sensitive data protection regulations such as the EU General Data Protection Regulation (GDPR), the California Consumer Privacy Act (CCPA)[2], and the People's Republic of China Multi-level Protection of Information Security (MLPS 2.0)[3]. On these grounds, Kavin and Ganapathy [10] propose a model for preserving data privacy and security on cloud-based systems. They claim that the utilisation of the new Chinese Remainder Theorem as a storage mechanism results in providing access only to authenticated users regarding cloud data. Further, Viejo and Sánchez [11] introduce a solution on IoT devices orchestration in order minimise security and privacy issues. The authors underline that, on the technical side, such a solution addresses the challenge of non-centralised data stream to fulfil the *data minimisation* principle included in the EU GDPR. In parallel, [12] suggests a *privacy-by-design* framework to address the design challenges of an IoT system due to its heterogeneous and distributed nature. They claim that their results conform to the GDPR directives as well as to data protection goals such as availability and confidentiality.

## 2.3. Capability-based access control

To provide a high level of security assurance and advanced data protection, the applications can generate a capability (i.e. an access token), which grants and provides access to the protected service/resource. On these grounds, the secured claims are mostly expressed by the Concise Binary Object Representation (CBOR) [35] and utilise CBOR Web Tokens (CWT), which are derived from a JSON Web-Token (JWT) for secure and stateless authorisations [36]. This approach enables the access control systems to manage the creation, the delegation and the revocation of the tokens. CWTs can be used as authentication credentials to control access in IoT systems that use low-power technologies.

In [37], the authors describe the Capability-Based Access Control system (CapBAC) which utilises the Policy Decision Point (PDP) to process an access request. The system retrieves the relevant access control policies from the policy repository and allows the Policy Enforcement Point (PEP) to delegate/enforce the access decisions.

The distributed nature of the Cloud-IoT solutions often requires that the certificate management, authentication and authorisation processes are independent. To further support complex resource management extensions, it is mandatory to define improved authorisation capabilities. Some of the objective grading criteria which can be used to improve the quality of the offered services are [38,39]: *(a)* Secure data distribution; *(b)* Efficient monitoring of the services; *(c)* Optimal resource utilisation. A recent work [40] also suggests that blockchain can be used to handle authentication, authorisation, and data access control token validation in order to follow the necessary GDPR policies.

## 2.4. Containerised microservices

Microservices consist an approach to systems architecture that builds on the software concept of modularisation. Several benefits are associated with microservices. Three of the most important ones are *faster delivery*, improved *scalability* and greater *autonomy*. Moreover, microservices rely on lightweight technologies (e.g. REST and HTTP) [41]. Time-critical industry applications, such as disaster early warning systems and live event broadcasting, often involve distributed components and intensive data communication and may include sensors in various geographical locations. Notwithstanding the cloud ecosystem's capability to support such applications, there is a need on deploying *resilient* and *reconfigurable applications* using the virtual runtime environment [42]. Nowadays, microservices can be packaged into containers, as lightweight virtualisation compared to virtual machines (VMs). Containers utilisation environments provide reactive auto-scaling methods which result to improvements of the application response time as well as of the resource pooling aspects [43].

## 3. The CAPODAZ architecture

Taking into consideration the related research work, we propose an innovative containerised authorisation and policy-driven architecture for facilitating security controls to the Cloud-IoT actors. This approach addresses the need to efficiently manage the access control services and to meet the underlying strict performance requirements. It also ensures that only the data that are strictly necessary to fulfil the service is revealed to the involved entities, thus fulfilling the *data minimisation* principle included in the *EU General Data Protection Regulation (GDPR)*.

We initiated this research to broaden the scope and performance of our previously work SeMMA [19]. SeMMA has been proposed as a general-purpose framework based on policy-aware security-enabled Service Oriented Computing (SOC) principles, which ensures a robust security management by deploying policy-driven security controls. That work aimed to study on the integration's methodological approach of secure M2M communications. The current work extends the original SeMMA capabilities and proposes a state-of-the-art architecture by utilising the cutting-edge paradigm of containerised microservices for several components of the proposed architecture and by incorporating real-life scenario components which apply within the Cloud-IoT ecosystem.

---

[2] https://oag.ca.gov/privacy/ccpa
[3] http://www.mps.gov.cn/n2254536/n4904355/c6159136/content.html

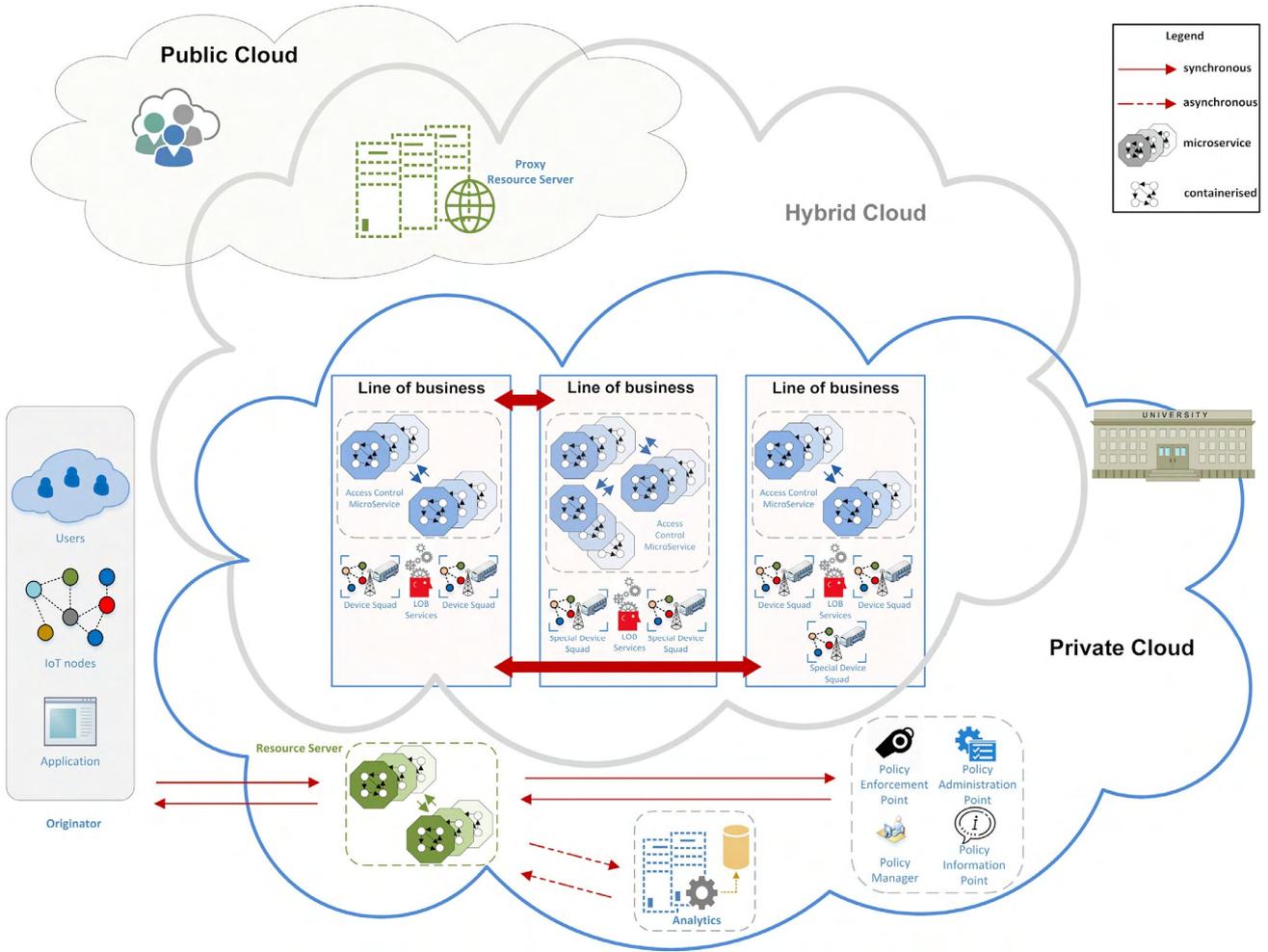

**Fig. 1.** The proposed CAPODAZ.

### 3.1. Proposed architecture

The proposed containerised authorisation and policy-driven architecture (CAPODAZ) provides a series of security enhancements for services which are offered within the Cloud-IoT ecosystem by incorporating microservices that lean upon Docker container clusters. CAPODAZ (Fig. 1) is an inclusive policy-based management access control architecture that encapsulates the policy-based and integration capabilities of SeMMA.

CAPODAZ leverages microservices, which can rapidly adapt to the operational challenges, adopt lean and software delivery, and facilitate more responsive and nimble software applications. Furthermore, the microservice schema can effectively control and manage the M2M-IoT actors and improve the Cloud-IoT service efficiency by maintaining stateless authentication assertions. These microservices are hosted on containers, which can maximise the scalability of the cloud infrastructure. Containers can also improve service partitioning and performance.

Due to the above, an interesting feature of the proposed architecture is that it supports and can adapt to various environments (e.g. enterprise, community).

The architecture also manages the multilevel integration of authentication and authorisation modules based on formal policy-based methods. Access control mechanisms are also provided for secure access to the Cloud-IoT resources. Table 1 illustrates the key elements and entities in CAPODAZ.

### 3.2. Policy-driven authorisations functional description

Fig. 2 illustrates the interactions between the entities in the policy-driven authorisations.

The *Originator* (i.e. an IoT actor) needs to communicate over constrained networks to send an access request to the *Resource Server*, which provides the protected resources and receives the requests from the client (i.e. the application). A capability token is generated and exchanged with the *Originator* to get access to the protected resources or the restricted service. The *Resource Server* utilises the authorisation services to validate the capability token and determine whether to process or deny the request. The *Resource Server* may utilise an external authorisation server for the verification of the capability tokens (i.e. a database lookup in the token table) and uses the dynamically established keys to protect the resources. The authorisation services interact with the *Registrar*, which keeps the created capability tokens and they provide the successful/error capability token response to the client with a set of claims/context/attribute values to the resource server about the authorisation. For instance, the authorisation services need to verify the validity of the capability token and, therefore, the access rights can be exercised. The keys need to be provisioned for generating and verifying the tokens with the resource server in advance, whereas the client needs to be trusted by the registered services before initiating the request. Additionally, the revocation of the tokens ensures that there are no stale requests. This capability is achieved by adding a lifetime to the tokens.

**Table 1**
Key elements and entities in CAPODAZ.

| Element/Entity | Role |
| --- | --- |
| Private cloud | Hosts any element of the CAPODAZ architecture apart from the resource proxy |
| Public cloud | Hosts several of the Cloud-IoT service proxy interfaces |
| Hybrid cloud | Hosts the core of CAPODAZ, i.e. the access control microservice |
| Line of business (LOB) | Term for referring to a product or a set of products; *here:* any IoT service |
| Originator | Term for referring to an IoT actor or an end-user application that initiates the requests to the resource server |
| Resource server | Provides the protected resources and receives the requests from the clients |
| Proxy resource server | Evaluates a public cloud user request to simplify and control the incoming load |
| Client | Represents an application to gain access to the protected resources |
| Access control microservice | Exposes an authorisation service. It logically hosts the Registrar and the authorisation services |
| Registrar | Stores the created capability tokens and responds to token authentication challenges by the Resource Server |
| Authorisation services | Negotiates access to a resource and verifies the Originators access rights |
| Policy information point (PIP) | Acts as a source of attribute values needed for authorisation policies |
| Policy administration point (PAP) | Includes a policy store (a repository with the domain policies) and provides the authoring and maintenance of the policy sets |
| Policy enforcement point (PEP) | Intercepts the resource access requests |
| Policy decision point (PDP) | Interacts with PIP to retrieve the relevant authorization policies and attributes and then evaluates the access request providing an authorization decision outcome to PEP |
| Analytics | Holds a cloud- and a service-enabled test suite and exposes a monolithic service for performance metering |
| Device squad | Term for referring to building blocks of any IoT service |
| Special device squad | Term for referring to special blocks of any IoT service |
| LOB service | Term for referring to any IoT service element |
| Synchronous | The request/response style of comms; the client initiates and waits for response |
| Asynchronous | The event-based style of comms; the client acknowledges other parties about what happened and expects them to know what to do |

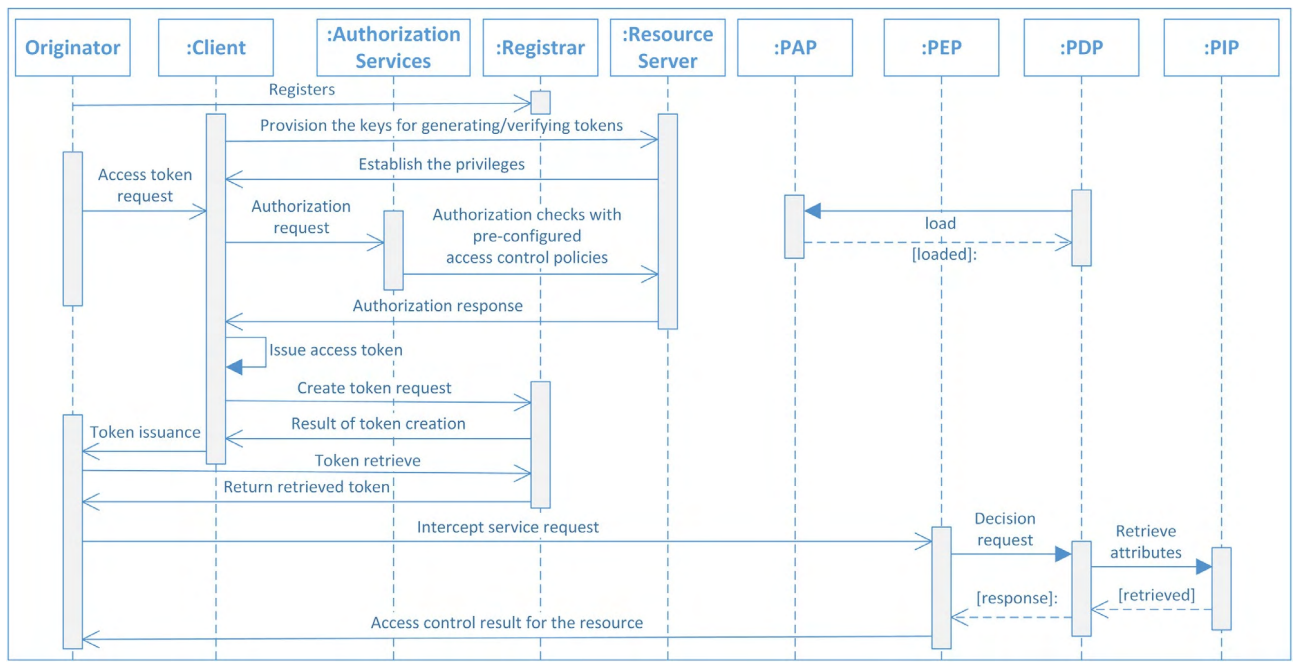

**Fig. 2.** Interaction of core entities in CAPODAZ.

In order to facilitate and to resolve authorisation policies at runtime, the capability token is sent to the API endpoint and calls the relevant policy method. Then, *PEP* intercepts the access request to determine whether to entitle and authorise the client to the resource. *PDP* evaluates the relevant policies so as to accept or reject the request. Upon receiving all the relevant attributes from *PIP* along with the access control policies from the policy store, *PDP* generates the policy decision. Provided that all the authorisation checks (i.e. subscription, payment-based and platform validation checks) have been completed, *PDP* updates *PEP* with the policy decision granting access to the protected resource based on short-lived capability tokens issued by the token authority for a finite amount of time and without sharing the originator's credentials. Access can be granted temporarily given the related attributes, the conditions and the originator's needs. Finally, *PEP* enforces the appropriate authorisation attributes decisions and delegates the access decision.

### 3.3. Added-value: containerised microservices with a cloud-based service provision

The containerised nature of the microservices deployment makes various innovative features possible. In contrast to monolithic applications, which contain domain-specific functionality and are packaged onto hardware pre-scaled for peak loads, the microservices can scale independently and almost instantaneously. The intra-service communication is handled through *pub-*

```
JWT response
{
 "aud": "Vehicle01",
 "user_name": "v01",
 "scope": ["read","trust"],
 "exp": 1518074605,
 "authorities": "ROLE_USER",
 "jti": "1d3b890201",
 "client_id": "CAPODAZ-client"
}
```

**Fig. 3.** The JWT registered claims-set of a CAPODAZ request.

*lish/subscribe* messages using an *event channel*, which carries granular portions of data.

Additionally, the *containerised* set-up of the proposed architecture provides an isolated, fully controlled and portable environment. A fine-grained systemic approach is followed here, which provides better isolation and, thus, service component cohabitation. The containerised nature of the proposed architecture maximises the benefits of the cloud-based scalable and elastic environment. Finally, the optimisation of time consumption and network utilisation issues in the execution mode are also crucial factors; these features are thoroughly discussed in Section 4.

## 4. Evaluation and results

For evaluating CAPODAZ, we utilise an IoT service prototype named iBuC [21,22], as a real-world scenario and a testbed. The iBuc prototype is an intelligent transportation service which runs in a University campus. For this prototype to offer a service with IoT-enabled features, multiple sensors and actuators are also incorporated regarding the autonomous vehicles' equipment, as well as the bus stations and the road infrastructure. For evaluating CAPODAZ, the Cloud-IoT actors are the sensors, the actuators, and the vehicles as well as the potential passengers who place service bookings using their smart/mobile device.

Furthermore, the access control service and the resource server of CAPODAZ are deployed on containerised microservices, as shown in Fig. 1. The policy service registry is XML-based, while the interaction between the policy engine entities conforms representations in XACML format. For the testing scenario, the line of business (LOB) service module (Table 1) of CAPODAZ can be a person/vehicle's localisation feature and the device squads can be autonomous vehicles or a network gateway within the University campus.

The following sections describe the apparatus set up, the methodology followed, the token exchange process and depict the evaluation results.

### 4.1. Apparatus

In the context of this work, several machine instances are deployed on a private as well as on a public cloud infrastructure-as-a-service (IaaS). The private cloud infrastructure hosts all the entities of the policy engine as well as the service endpoints of the iBuC prototype for the real-life scenario testing. It also accommodates the access control core module of CAPODAZ. The public cloud hosts a proxy resource server to receive inter-domain requests and to forward them to its private cloud counterpart (i.e. the resource server) in a simplified and secure manner.

Regarding the RESTful microservices setup, we utilise four similar frameworks, namely: CAPODAZ, *Spark* [44], *Wildfly-Swarm* [45] and *Spring-boot-Tomcat* [46]. The proposed CAPODAZ prototype development is initially based on *light-4j* [47] and JAX-RS 2.0 lightweight Java.

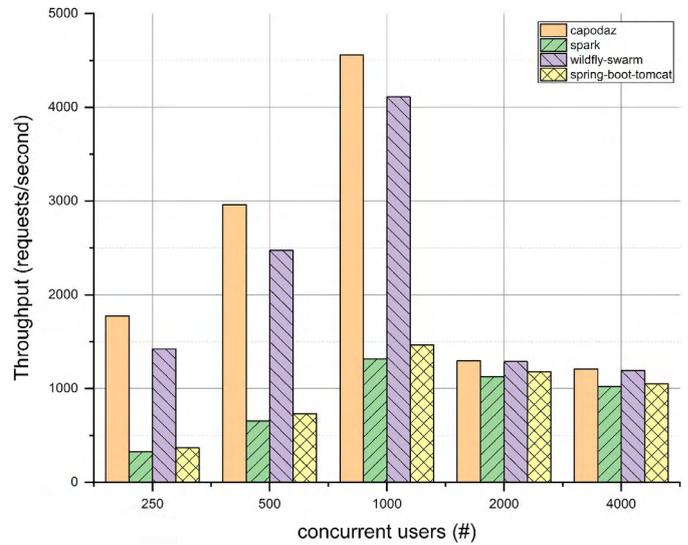

**Fig. 4.** CWT requests for an authorisation.

**Fig. 5.** Throughput of successful request (binomial load).

*Docker v18.06.1-ce* is utilised for the microservice containerisation and the clusters' (swarms) deployment; *Keycloak v4.8.0* [48] is used as the identity and access management (IAM) service; *Ubuntu LTS Server v16.04* is the operating system for all the deployments. The proposed architecture's *Analytics* entity, which is deployed as a real-time benchmark tool, exposes a monolithic metering service using *Apache Jmeter v5.0*, *Python v3.0* and *Java v8*. Table 2 summarises all the entities' information for the testing scenario.

**Table 2**
Summary of the apparatus setup.

| Cloud instance IP address | Role | Type of service | OS, frameworks and libraries | port(s) | Hardware specs |
|---|---|---|---|---|---|
| 83.xxx.xxx.xx4 | Analytics | Monolithic | Ubuntu LTS, Apache Web Server, Apache JMeter Python, Java, | 80 | vCPU=QEMU AMD (64-bit, 8 cores) // Mem. = 8GB // HD=40GB |
| 83.xxx.xxx.xx1 83.xxx.xxx.xx9 83.xxx.xxx.xx6 83.xxx.xxx.xx8 83.xxx.xxx.xx2 83.xxx.xxx.xx3 | Container Cluster (16x each) // Resource Server // AAA | Microservice | Ubuntu LTS, Spark, Wildfly-swarm, Spring-boot-Tomcat, CAPODAZ, Keycloak | 8082 ~ 8086 | |
| 195.xxx.xxx.xx5 | Proxy Resource Server | Monolithic | Ubuntu LTS, Apache Web Server | 80 | |

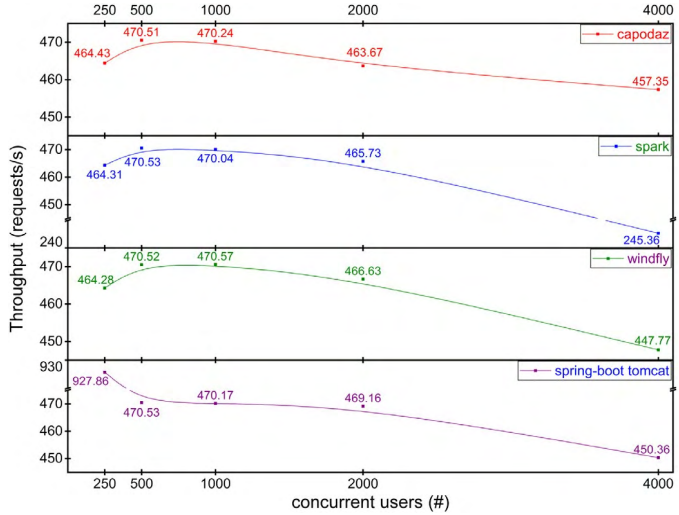

**Fig. 6.** Service throughput (uniform load).

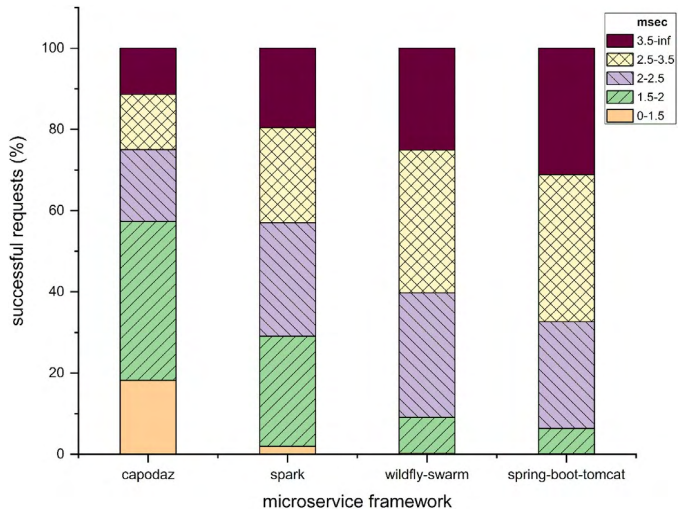

**Fig. 7.** Time burstiness on successful requests (uniform load).

### 4.2. Methodology

*Three-rounds evaluation process.* To evaluate the proposed CAPO-DAZ architecture, we go through a 3-rounds process and compare it with three similar microservice frameworks focusing on the authorisation process. Since CAPODAZ is implemented on a cloud-based infrastructure, we choose to utilise a tailored cloud-based monolithic system to initiate and to control the benchmarking process for various user populations (i.e. 250, 500, 1000, 2000 and 4000) which access the four microservice frameworks to simulate IoT-actors interaction with the proposed architecture. To eliminate network delays and bottlenecks as well as software misconfigurations, this testing infrastructure is developed at the private cloud infrastructure on which the tested microservice frameworks are deployed.

**Table 3**
Load patterns for the microservices.

| Load CR (distribution) | IoT-actors CR (concurrent users) | Parameters |
|---|---|---|
| Binomial | {250, 500, 1000, 2000, 4000} | 1000 reqs/s, CR no delays between requests |
| Uniform | {250, 500, 1000, 2000, 4000} | 1s delay CR for each user between requests |
| Poisson | {1000} | $\lambda = (1, 0.2, 0.0022, 0.0011)$ |

The 3-rounds testing process is as follows:

(a) the first round comprises a test for the successful requests for a population of virtual users, given the fact that a binomial-distributed load towards each microservice exists;
(b) the second round illustrates a uniform-distributed test for the same user population and time in order to elect the set(s) of users that the microservice framework can handle their requests in an acceptable rate;
(c) the third round presents a comparison of the uniform distributed requests to a Poisson distribution of requests for the elected set of 1000 users.

*Experimental time.* The experimental time period for each setup is 15min; other testing periods (i.e. 30min, 1h, 12h) have also been tested and we have observed that the variances of the results are not higher than 3–5%. Hence, those results fall outside the scope of this work. However, in most cases, the 15 min-experiments give a maximum of 0.5 million requests, which is a good sample for calculations and discussion.

*Load.* Table 3 depicts how the load patterns (i.e. # of events) are statistically distributed [49–51].

*Measurement metrics.* We evaluate *latency*, *throughput* and *successful requests*.

*Authorisation setup.* In CAPODAZ, the access requests flow through the decoupled microservices. To implement the proposed distributed capability-based access control approach, we use CBOR to represent the capability token because of its suitability in constrained environments, such as those that may exist in the iBuC scenario. CWTs support built-in expiry mechanisms, designed for small code and message size and they can be used in the responses between the client and the Resource Server enabling fine-grained access control information within the token. There is a twofold goal in this process: first, the calling clients must gain access to the endpoints and to the services according to the policies, and, second, the successful authorisation which allows the call to process the request.

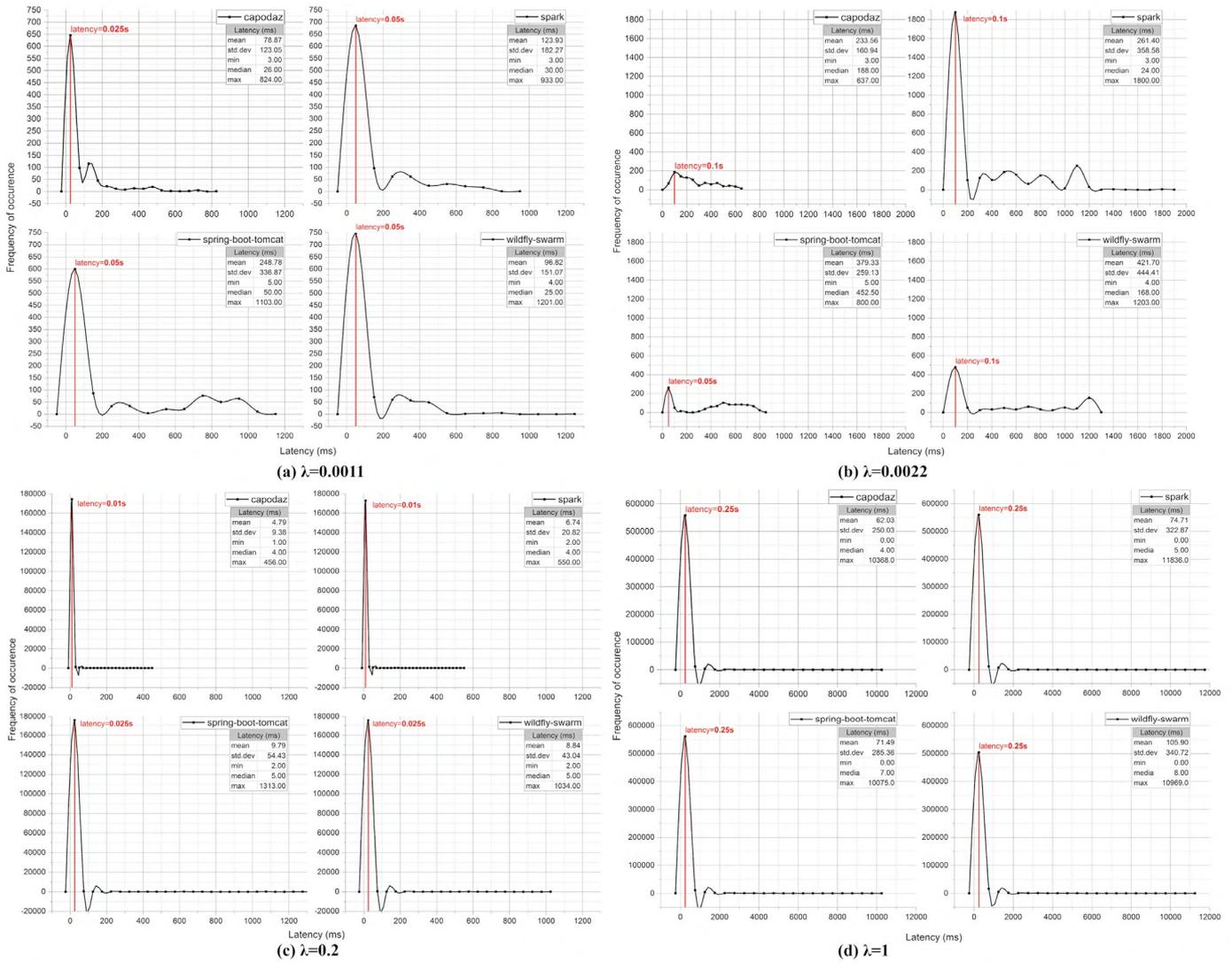

**Fig. 8.** Fluctuation of latency (Poisson load).

### 4.3. Tokens exchange

*Java Web Token (JWT).* As mentioned in Section 3, CAPODAZ has inherited from SeMMA policy-based security capabilities. As such, CAPODAZ supports the password, the symmetric-key and the token as grant types during the authorisation process. Apart from these grant types, the use of HTTP basic authentication, creates a key and a capability token during the generation of a new API in order to protect it. After sending the access request, the authorisation of the capability token is granted by the authorisation server. If the application identity is authenticated and the authorisation grant is valid, then the capability token is returned in response. The sample payload of the JWT is shown in Fig. 3.

*CBOR Web Tokens (CWT).* In the case of CWT, the client credentials grant uses client-id and client secret in the request payload. Nonetheless, the scheme can be further extended with certificates or Datagram Transport Layer Security (DTLS) pre-shared keys. Selecting the proper grant type depends upon the use case and the specific application type given that the client and the resource server support the proper security encoding and attributes. Other parameters weigh in as well, such as the level of trust for the application, and service type information. In Fig. 4, the payloads for diverse grant types using the REST APIs are presented: first, the response containing an access token bound to symmetric key where transport layer security is with CBOR encoding; second, the token request and response using client credentials with CBOR encoding; last, the token refresh grant type grant where COSE is used to provide object-security.

### 4.4. Results

#### 4.4.1. First round: binomial-distributed load

We use binomial distributed requests to evaluate the probability of "success" and "not success" requests. Table 4 depicts this round's results and Fig. 5 illustrates the output graph.

Remarkably, CAPODAZ and Wildfly-Swarm have a 100% rate of successful requests for the case of *users = {250, 500, 1000}* and they reach the best response rate for all sets of users. Although CAPODAZ's response rate considerably decreases for the user-sets of 2000 and 4000, it still remains at the first place among the four frameworks.

**Table 4**
Binomial distributed load.

| Concurrent users | Throughput (requests/s) | | | |
|---|---|---|---|---|
| | capodaz | spark | wildfly-swarm | spring-boot-tomcat |
| 250 | 1775 | 327 | 1423 | 367 |
| 500 | 2961 | 653 | 2473 | 733 |
| 1000 | 4558 | 1316 | 4111 | 1463 |
| 2000 | 1296 | 1127 | 1288 | 1180 |
| 4000 | 1207 | 1020 | 1192 | 1051 |

*4.4.2. Second round: uniform-distributed load*

We use a uniformly distributed load to evaluate the frameworks, assuming equal arrival probabilities for each event. Fig. 6 highlights the service throughput for different set of users and Table 5 summarises the results of this round.

These tests revealed (Fig. 6) that, apart from Spring-boot-Tomcat which showed a bursty throughput on the small number of users (i.e. 250), there is a notable throughput increase for the rest of the four frameworks up to the level of 1000 users. On average, CAPODAZ, Wildfly-Swarm and Spark resulted in barely distinguishable values for the sets of *users = {250, 500, 1000}*. Further, the gradient of the throughput curve is interestingly similar for the Spark, Wildfly-Swarm and Spring-boot-Tomcat frameworks. In the same vein, we found that CAPODAZ has the highest throughput for 4000 users. Nevertheless, the service throughput for all four frameworks is considerably lower when compared to the first-round results.

Fig. 7 shows the timeframe in which the microservice frameworks respond successfully.

A notable remark (Fig. 7) about CAPODAZ is that it achieves about 60% of its successes within the first 2 ms, while, in the same timeframe, the other frameworks achieve about: 30% (Spark), 10% (Wildfly-Swarm) and below 10% (Spring-boot-Tomcat). Compared to the first-round results with the binomial load, CAPODAZ also demonstrates its bursty nature for the uniform-distributed input load.

*4.4.3. Third round: Poisson-distributed load*

We use a Poisson arrival distribution to evaluate each framework's behaviour, while responding to an infinite number of event arrivals. We work on several lambda ($\lambda$) values for the input load and consequently observe each framework's behaviour. We focus on the set of *1000 users* for three reasons:

(a) it is the median of the users' sets;
(b) as derived from the second round, all the microservice frameworks increase up to or reach the maximum of their throughput at the set of 1000 users;
(c) this user set is realistic enough for the iBuC real-life scenario (i.e. not more than 1000 Cloud-IoT actors are interacting at the same time within the University campus).

As anticipated, the findings of the third round prove that CAPODAZ has the lowest mean latency and the lowest max latency values for any given lambda ($\lambda$). It is also demonstrated that the proposed microservice architecture has the best behaviour = regarding the standard deviation and most appearances of latency values when $\lambda = 0.2$ and $\lambda=0.0011$ (i.e. 0.01s and 0.025s respectively). Nevertheless, when the input load is distributed by = *1*, then CAPODAZ shows the highest throughput (i.e. *469.85 req/s*). These findings are presented in Table 6 and Fig. 8.

As demonstrated in Fig. 9, CAPODAZ has the best mean latency value for the set of *users = {1000}* which have uniformly distributed requests. Moreover, its maximum latency value is the best when compared to the tested frameworks. Finally, the majority of requests are fulfilled by CAPODAZ with a latency value of *0.25s*, an observation that highlights its bursty nature.

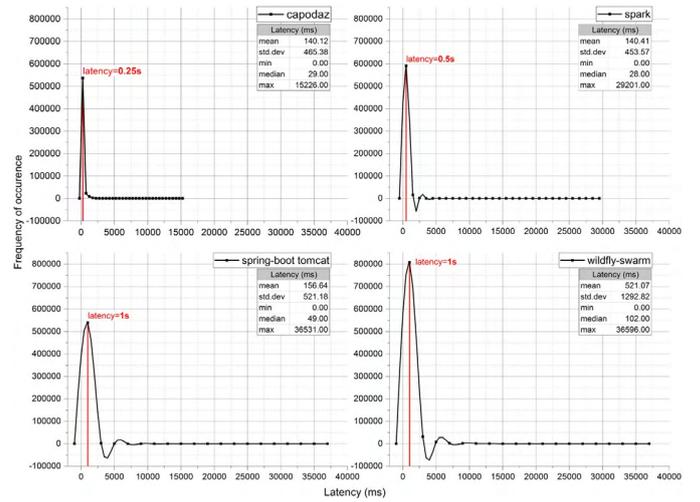

**Fig. 9.** Fluctuation of latency (uniform load).

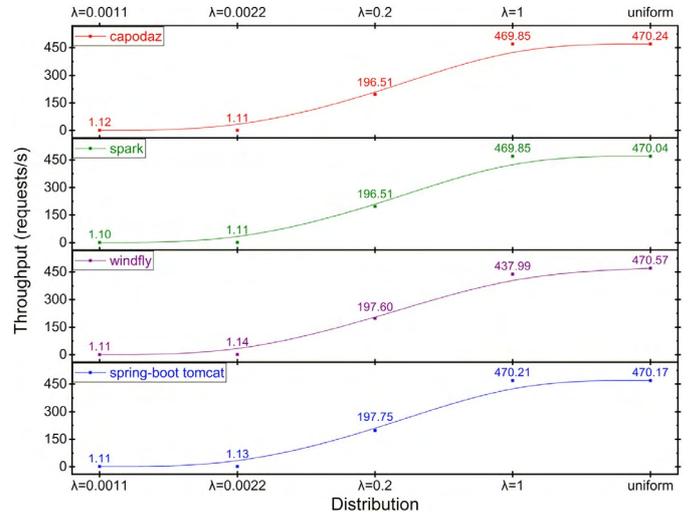

**Fig. 10.** Service throughput (1000 users: Poisson vs uniform load).

As stated, Table 6 and Figs. 8 and 9 depict that CAPODAZ achieves the lower latency values in any Poisson-distributed or uniform-distributed load. With respect to the mean latency values, CAPODAZ is also the best among the tested microservice frameworks. Fig. 10 shows that the CAPODAZ = throughput is the highest when $\lambda=0.0011$ and the lowest when $\lambda=0.2$. However, even in the latter case, the CAPODAZ throughput does not significantly deviate.

In conclusion, we evaluate the CAPODAZ framework to three similar microservice approaches. In the first round, the frameworks were tested upon a binomial-distributed load and CAPODAZ shown the best response rate. In the second round, upon a uniform-distributed load, CAPODAZ achieved about 60% of the successive requests in just 2ms and shown the highest burstiness. In the third round, upon the more life-realistic Poisson-distributed load, CAPODAZ had the lowest latency and the highest throughput (i.e. when the requests were arriving in a high rate, $\lambda=0.0011$).

*4.5. Other critical performance factors*

As discussed in the previous sections, the execution performance of the event policies depends on the computing machinery hardware characteristics. In the case of event scripting and ap-

**Table 5**
Latency and throughput of successful requests for a uniform-distributed input load.

| concurrent users | u250 | | | | u500 | | | | u1000 | | | | u2000 | | | | u4000 | | | |
|---|---|---|---|---|---|---|---|---|---|---|---|---|---|---|---|---|---|---|---|---|
| | fr1 | fr2 | fr3 | fr4 | fr1 | fr2 | fr3 | fr4 | fr1 | fr2 | fr3 | fr4 | fr1 | fr2 | fr3 | fr4 | fr1 | fr2 | fr3 | fr4 |
| **Mean Latency (ms)** | 25.09 | 26.89 | 27.82 | 27.59 | 56.08 | 62.73 | 54.94 | 62.04 | 140.12 | 140.41 | 157.68 | 156.64 | 470.81 | 489.63 | 439.72 | 506.25 | 1237.93 | 874.36 | 809.33 | 720.48 |
| **Throughput (requests/s)** | 464.43 | 464.31 | 464.28 | 927.86 | 470.51 | 470.53 | 470.52 | 470.53 | 470.24 | 470.04 | 470.57 | 470.17 | 463.67 | 465.73 | 466.63 | 469.16 | 457.35 | 245.36 | 447.77 | 450.36 |
| **Load (KB/s)** | 0.04 | 0.05 | 0.04 | 0.03 | 0.07 | 0.10 | 0.07 | 0.07 | 0.14 | 0.19 | 0.15 | 0.14 | 0.29 | 0.38 | 0.30 | 0.27 | 0.57 | 0.76 | 0.60 | 0.55 |

For space reasons, *fr1…fr4* stands for capodaz, spark, wildfly-swarm and spring-boot-tomcat respectively.

**Table 6**
Latency and throughput of successful requests for 1000 users (Poisson and uniform load).

| | $\lambda = 0.0011$ | | | | $\lambda = 0.0022$ | | | | $\lambda = 0.2$ | | | | $\lambda = 1$ | | | | uniform | | | |
|---|---|---|---|---|---|---|---|---|---|---|---|---|---|---|---|---|---|---|---|---|
| | fr1 | fr2 | fr3 | fr4 | fr1 | fr2 | fr3 | fr4 | fr1 | fr2 | fr3 | fr4 | fr1 | fr2 | fr3 | fr4 | fr1 | fr2 | fr3 | fr4 |
| **Mean latency (ms)** | 78.87 | 123.93 | 96.82 | 248.78 | 233.56 | 261.40 | 421.70 | 379.33 | 4.79 | 6.74 | 8.84 | 9.79 | 62.03 | 74.71 | 105.9 | 71.49 | 140.12 | 140.41 | 521.07 | 156.64 |
| **Most appearances (s)** | 0.025 | 0.05 | 0.05 | 0.05 | 0.1 | 0.1 | 0.1 | 0.05 | 0.01 | 0.01 | 0.025 | 0.025 | 0.25 | 0.25 | 0.25 | 0.25 | 0.25 | 0.5 | 1 | 1 |
| **Std. dev** | 123.05 | 182.27 | 151.07 | 336.87 | 160.94 | 358.58 | 444.41 | 259.13 | 9.38 | 20.82 | 43.04 | 54.43 | 250.03 | 322.87 | 340.72 | 285.36 | 456.38 | 453.57 | 1292.82 | 521.18 |
| **Throughput (requests/s)** | 1.12 | 1.10 | 1.11 | 1.11 | 1.11 | 1.11 | 1.14 | 1.13 | 196.51 | 196.09 | 197.60 | 197.75 | 469.85 | 469.59 | 437.99 | 470.21 | 470.24 | 470.04 | 470.57 | 470.17 |
| **Load (KB/s)** | 0.14 | 0.19 | 0.15 | 0.15 | 0.14 | 0.19 | 0.15 | 0.14 | 0.14 | 0.19 | 0.15 | 0.14 | 0.14 | 0.19 | 0.15 | 0.14 | 0.14 | 0.19 | 0.15 | 0.14 |

For space reasons, *fr1…fr4* stands for capodaz, spark, wildfly-swarm and spring-boot-tomcat respectively.

**Table 7**
Other critical factors for policy-based management.

| Policy-metrics | Factors |
|---|---|
| Policy definition | Complexity |
| | Detection of updates difficult to update |
| | Vulnerabilities |
| | Outdates rules |
| Policy execution | Load/retrieval |
| | Performance |
| | Difficult to manage |
| | Frequency of resource request |
| | Difficult to measure |
| Policy enforcement | Risks |
| | Urgency |
| Run time policy validation | Accuracy |
| | Correctness |
| | Performance |
| | Security violations |
| | Flaws |
| | Data quality |

plets, the performance depends on the *service platform*, the *load-level* and the *total amount of available memory*. To overcome some of these performance limitations, CAPODAZ is proposed as a containerised cloud-based architecture, in order to maximise the benefits of the cloud-based scalable and elastic environment. In the same vein, CAPODAZ also scrutinises the microservice paradigm to increase performance in execution mode.

Nevertheless, other performance factors still exist. Table 7 summarises these critical factors which can affect any proposed policy-based management solution.

## 5. Conclusions and future work

In this work, we proposed a containerised authorisation and policy-driven architecture (CAPODAZ) using microservices in a cloud-based environment. CAPODAZ encapsulates our previous work on the SeMMA architectural features and it is integrated in the iBuC intelligent transportation service prototype. We tested CAPODAZ using a capability token-based scheme to facilitate the authorisations in a real-life IoT scenario IoT for intelligent transportation services. Then, we evaluated the proposed architecture against similar microservice frameworks and focused on the *load-level* factor; that is on, latency, throughput and successful requests. We utilised the Docker built-in engine, which relies on the Linux operating system's processes synchronous communication, for the intra-microservice messaging service. CAPODAZ was shown as the most suitable framework for the set of 1,000 users, which is an appropriate population size for the tested real-life scenario. The best performance of CAPODAZ was noted in the case of the most realistic situation (i.e. Poisson), where the load was unevenly time-distributed and the requests were arriving in a high pace. The proposed microservice architecture can be deployed on any cloud-based environment due to its containerised nature.

Our future work includes experimentation with other event-based messaging services (such as Kafka, and RabbitMQ) [52] and with different distributions (e.g. Zipf's Law) regarding the service originators (i.e. IoT-actors). We also intend to conduct further analysis, modelling and implementation to evaluate architectural and performance aspects when using a centralised or a distributed policy engine in resource constrained environments [53].